\documentstyle[multicol,aps,psfig]{revtex}
\renewcommand{\narrowtext}{\begin{multicols}{2}
\global\columnwidth20.5pc\noindent}
\renewcommand{\widetext}{\end{multicols}
\global\columnwidth42.5pc}
\begin{document}
\draft
\preprint{13 November 2003}
\title{Nuclear Magnetic Relaxation in the Haldane-Gap Antiferromagnet
       Ni(C$_2$H$_8$N$_2$)$_2$NO$_2$(ClO$_4$)}
\author{Shoji Yamamoto and Hiromitsu Hori}
\address{Division of Physics, Hokkaido University,
         Sapporo 060-0810, Japan}
%\date{Received \hspace{4cm}}
\date{Received 13 November 2003}
\maketitle
\begin{abstract}
A new theory is proposed to interpret nuclear spin-lattice relaxation-time
($T_1$) measurements on the spin-$1$ quasi-one-dimensional Heisenberg
antiferromagnet Ni(C$_2$H$_8$N$_2$)$_2$NO$_2$(ClO$_4$) (NENP).
While Sagi and Affleck pioneeringly discussed this subject in terms of
field-theoretical languages, there is no theoretical attempt yet to
explicitly simulate the novel observations of $T_1^{-1}$ reported
by Fujiwara {\it et al.}.
By means of modified spin waves, we solve {\it the minimum of $T_1^{-1}$
as a function of an applied field}, pending for the past decade.
\end{abstract}
% 75.10.Jm: Quantized spin models
% 76.50.$+$g: Ferromagnetic, antiferromagnetic, and ferrimagnetic
% 05.30.Jp: Boson systems
% 75.40.Mg: Numerical simulation studies
%             resonances; spin-wave resonance
% 75.50.Xx: Molecular magnets
% 76.60.$-$k: Nuclear magnetic resonance and relaxation
\narrowtext

   Predicting a striking contrast between integer- and
half-odd-integer-spin one-dimensional Heisenberg antiferromagnets, Haldane
\cite{H464,H1153} sparked renewed interest in low-dimensional quantum
magnetism.
The Haldane gap, that is, a magnetic excitation gap immediately above the
ground state, was not only calculated by various numerical tools
\cite{W3844,G3037,Y3348,S493,W14529,T047203} but indeed observed in
spin-$1$ quasi-one-dimensional Heisenberg antiferromagnets such as
CsNiCl$_3$ \cite{B371} and Ni(C$_2$H$_8$N$_2$)$_2$NO$_2$(ClO$_4$) (NENP)
\cite{R945}.
The valence-bond-solid model \cite{A799,A477} significantly contributed
toward understanding novel features of the Haldane massive phase such as
fractional spins induced on the boundaries \cite{W2863,M913}, a string
order hidden in the ground state \cite{N4709,Y9528} and magnon excitations
against the hidden order \cite{K627,F8983,Y157,Y1795}.
The nonlinear-$\sigma$-model quantum field theory \cite{A397,A409}
skillfully visualized the competition between massive and massless phases,
while a generalized Lieb-Schultz-Mattis theorem \cite{O1984,T103} gave a
criterion for the gap formation in a magnetic field.

   Recent progress in the experimental studies also deserves special
mention.
Not only the single-magnon dispersion relation \cite{T2313} but also the
two-magnon continuum \cite{W3844,Y545} was directly observed by
inelastic-neutron-scattering measurements on NENP \cite{M3571} and
CsNiCl$_3$ \cite{Z017202}.
An applied magnetic field may destroy the Haldane gap and bring back
magnetism to the system.
Such a field-induced long-range order was indeed realized in a nickel
compound Ni(C$_5$H$_{14}$N$_2$)$_2$N$_3$(PF$_6$) \cite{H2566}.
Another family of linear-chain nickelates of general formula
$R_2$BaNiO$_5$ ($R=\mbox{rare\ earth\ or\ Y}$) \cite{D409} exhibited
a novel scenario of one- to three-dimensional crossover \cite{M68,T15189}.
When the nonmagnetic Y$^{3+}$ ions are substituted by other magnetic
rare-earth ions in Y$_2$BaNiO$_5$ with a disordered ground state, there
appears a three-dimensional long-range order, while the one-dimensional
gapped excitations persist both above and below the N\'eel temperature
\cite{Z6437,Z7210,Y11516}.

   Nuclear spin-lattice relaxation time ($T_1$) has also been measured on
Haldane-gap antiferromagnets.
The field ($H$) dependence of $T_1^{-1}$ is of particular interest at both
high and low temperatures.
The high-temperature relaxation rate has been discussed in the context of
transport properties.
Takigawa {\it et al.} \cite{T2173} measured the high-temperature
relaxation rate of $^{31}$P and $^{51}$V nuclei of AgVP$_2$S$_6$, which is
also an ideal spin-$1$ Haldane-gap antiferromagnet, and observed the
diffusive dynamics $T_1^{-1}\propto H^{-1/2}$ \cite{B2215}.
There is a hot argument \cite{S943,F2810,S2712,F2714} whether the spin
transport in quantum spin-gapped antiferromagnets should be diffusive or
ballistic at finite temperatures.
On the other hand, Gaveau {\it et al.} \cite{G647} and Fujiwara
{\it et al.} \cite{F7837,F11860} measured the low-temperature relaxation
rate of $^1$H nuclei of NENP.
As an applied field increases, the first excited state moves down and then
crosses the ground-state energy level.
Indeed $T_1^{-1}$ reaches a peak near the critical field
$H_{\rm c}\equiv{\mit\Delta}_0/g\mu_{\rm B}$ at every temperature, but it
is not a monotonically increasing function of $H$, at low temperatures
$T\alt{\mit\Delta}_0/k_{\rm B}$ in particular.
Sagi and Affleck \cite{S9188} formulated the nuclear magnetic resonance in
Haldane-gap antiferromagnets in terms of field-theoretical languages.
However, few investigations have followed their pioneering argument and
there is no attempt yet to explicitly fit a theory for the above
observations.

   In such circumstances, we revisit the low-temperature nuclear magnetic
relaxation in Haldane-gap antiferromagnets with particular emphasis on its
field dependence.
Excluding any phenomenological assumption from our argument, we calculate
the nuclear spin-lattice relaxation rate by means of modified spin waves.
A new theory claims that {\it as an applied field increases, $T_1^{-1}$
should initially decrease logarithmically and then increase
exponentially}, well explaining experimental observations.

   We employ the spin-$1$ one-dimensional Heisenberg Hamiltonian
\begin{equation}
   {\cal H}
      =J\sum_{l=1}^L
        \mbox{\boldmath$S$}_{l} \cdot \mbox{\boldmath$S$}_{l+1}
      -g\mu_{\rm B}H\sum_{l=1}^L S_l^z\,.
   \label{E:H}
\end{equation}
In order to illuminate the essential relaxation mechanism in spin-gapped
antiferromagnets as analytically as possible, we do not take any
anisotropy into consideration in this letter.
Magnetic anisotropy quantitatively affects the gap amplitude but has no
qualitative effect on the whole scenario.
Scaling temperature and an applied field by the Haldane gap, we present a
universal theory.
Quantitative refinement of the final product in the presence of single-ion
and orthorhombic anisotropy terms will be considered elsewhere.
Introducing the Holstein-Primakoff bosonic operators
\begin{equation}
   \left.
   \begin{array}{lll}
      S_{2n-1}^+=\sqrt{2S-a_n^\dagger a_n}\ a_n\,,&
      S_{2n-1}^z=S-a_n^\dagger a_n\,,\\
      S_{2n}^+=b_n^\dagger\sqrt{2S-b_n^\dagger b_n}\ ,&
      S_{2n}^z=-S+b_n^\dagger b_n\,,
   \end{array}
   \right.
   \label{E:HP}
\end{equation}
and retaining only bilinear terms of them, we rewrite the Hamiltonian as
\begin{eqnarray}
   &&
   {\cal H}=-2JS^2N
   \nonumber \\
   &&\quad\ \ 
   +(2JS+g\mu_{\rm B}H)\sum_{n=1}^N a_n^\dagger a_n
   +(2JS-g\mu_{\rm B}H)\sum_{n=1}^N b_n^\dagger b_n
   \nonumber \\
   &&\quad\ \ 
   +JS\sum_{n=1}^N
    \left(
     a_n^\dagger b_n^\dagger+a_n b_n
    +b_n^\dagger a_{n+1}^\dagger+b_n a_{n+1}
    \right)\,,
   \label{E:HHP}
\end{eqnarray}
where $N\equiv L/2$.
In order to preserve the up-down symmetry, or the sublattice symmetry, we
optimize the spin-wave distribution functions constraining
the total staggered magnetization to be zero
\cite{T2494,H4769,T5000,Y769}:
\begin{equation}
   \sum_n\left(a_n^\dagger a_n+b_n^\dagger b_n\right)=2NS\,.
   \label{E:const}
\end{equation}
The constraint is enforced by introducing a Lagrange multiplier and
diagonalizing an effective Hamiltonian
\begin{equation}
   \widetilde{\cal H}
   ={\cal H}
   +2J\lambda\sum_n\left(a_n^\dagger a_n+b_n^\dagger b_n\right)\,.
   \label{E:effH}
\end{equation}
Via the Bogoliubov transformation
\begin{equation}
   \left.
   \begin{array}{lll}
      {\displaystyle\frac{1}{\sqrt{N}}}
      {\displaystyle\sum_n}
      {\rm e}^{ {\rm i}k(2n-1/2)}a_n
      &=&
      \alpha_k{\rm cosh}\theta_k-\beta_k^\dagger {\rm sinh}\theta_k\,,\\
      {\displaystyle\frac{1}{\sqrt{N}}}
      {\displaystyle\sum_n}
      {\rm e}^{-{\rm i}k(2n+1/2)}b_n
      &=&
      \beta_k {\rm cosh}\theta_k-\alpha_k^\dagger{\rm sinh}\theta_k\,,
   \end{array}
   \right.
   \label{E:FT}
\end{equation}
with ${\rm tanh}2\theta_k=S{\rm cos}k/(S+\lambda)$, we reach the spin-wave
Hamiltonian
\begin{eqnarray}
   &&
   {\cal H}=-2JS^2N-2J(S+\lambda)N+J\sum_k\omega_k
   \nonumber \\
   &&\qquad
   +J\sum_k\left(\omega_k^-\alpha_k^\dagger\alpha_k
                +\omega_k^+\beta_k^\dagger \beta_k\right)\,,
   \label{E:SWH}
\end{eqnarray}
where
\begin{equation}
   \omega_k^\pm \mp g\mu_{\rm B}H/J
   =2\sqrt{(S+\lambda)^2-S^2{\rm cos}^2k}\equiv \omega_k\,.
\end{equation}
Minimization of the free energy gives the optimum distribution functions
as $\bar{n}_k^\pm=({\rm e}^{J\omega_k^\pm/k_{\rm B}T}-1)^{-1}$ and
$\lambda$ is then determined through
\begin{equation}
   \sum_k\left(\bar{n}_k^- +\bar{n}_k^+ +1\right){\rm cosh}2\theta_k
   =(2S+1)N\,.
\end{equation}

   Now we calculate the relaxation rate in terms of the modified spin
waves.
Considering the electronic-nuclear energy-conservation requirement,
the Raman scattering predominates in spin-gapped antiferromagnets.
The Raman relaxation rate is generally given by
\begin{eqnarray}
   &&
   \frac{1}{T_1}
    =\frac{4\pi(g\mu_{\rm B}\hbar\gamma_{\rm N})^2}
          {\hbar\sum_i{\rm e}^{-E_i/k_{\rm B}T}}
     \sum_{i,j}{\rm e}^{-E_i/k_{\rm B}T}
   \nonumber \\
   &&\qquad\times
     \big|
      \langle j|{\scriptstyle\sum_l}A_lS_l^z|i\rangle
     \big|^2
     \,\delta(E_j-E_i-\hbar\omega_{\rm N})\,,
\label{E:T1def}
\end{eqnarray}
where $A_l$ is the dipolar coupling constants between the nuclear and
electronic spins in the $l$th site, $\omega_{\rm N}\equiv\gamma_{\rm N}H$
is the Larmor frequency of the nuclei with $\gamma_{\rm N}$ being the
gyromagnetic ratio, and the summation $\sum_i$ is taken over all the
electronic eigenstates $|i\rangle$ with energy $E_i$.
By means of the modified spin waves, eq. (\ref{E:T1def}) is rewritten as
\begin{eqnarray}
   &&
   \frac{1}{T_1}
    =\frac{4\pi(g\mu_{\rm B}\hbar\gamma_{\rm N})^2}
          {\hbar N^2\sum_i{\rm e}^{-E_i/k_{\rm B}T}}
     \sum_{i,j}{\rm e}^{-E_i/k_{\rm B}T}
     \delta(E_j-E_i-\hbar\omega_{\rm N})
   \nonumber \\
   &&\qquad\times
     \Bigl|
      \langle j|
      {\scriptstyle\sum_{k,k'}}A_{k'-k}
      \bigl[
       \big(
        \alpha_{k'}^\dagger\alpha_k-\beta_{k'}^\dagger\beta_k
       \big){\rm cosh}\theta_{k'}{\rm cosh}\theta_k
   \nonumber \\
   &&\qquad\quad
      -\big(
        \alpha_{k'}\alpha_k^\dagger-\beta_{k'}\beta_k^\dagger
       \big){\rm sinh}\theta_{k'}{\rm sinh}\theta_k
      \bigr]
      |i\rangle
     \Bigr|^2\,,
   \label{E:T1SW}
\end{eqnarray}
where $A_q=\sum_l{\rm e}^{{\rm i}ql}A_l$.
The Fourier components of the hyperfine coupling constant exhibit little
momentum dependence when the nuclei take unsymmetrical positions in the
crystal, which is the case with the protons in NENP \cite{F11860}.
Hence we assume in the following that $A_q\simeq A_{q=0}\equiv A$.
Due to the significant difference between the electronic and nuclear
energy scales ($\hbar\omega_{\rm N}\alt 10^{-5}J$), eq. (\ref{E:T1SW})
ends in
\begin{equation}
   \frac{1}{T_1}
   =\frac{4(g\mu_{\rm B}\hbar\gamma_{\rm N}A)^2}{\pi\hbar}
    \int_{-\pi/2}^{\pi/2}
    \frac{\sum_{\sigma=\pm}\bar{n}_k^\sigma(\bar{n}_k^\sigma+1)}
         {\sqrt{v^2k^2+v\hbar\omega_{\rm N}}}
    {\rm d}k\,,
   \label{E:T1}
\end{equation}
where assuming moderate temperatures
$k_{\rm B}T\ll\omega_{k=\pi/2}^-=2J(S+\lambda)-g\mu_{\rm B}H$, we have
approximated the dispersion relations as
\begin{equation}
   J\omega_k^\pm\simeq{\mit\Delta}+vk^2\pm g\mu_{\rm B}H\,,
\end{equation}
with
\begin{equation}
   {\mit\Delta}=2J\sqrt{\lambda(2S+\lambda)}\,,\ \ 
   v=\frac{JS^2}{\sqrt{\lambda(2S+\lambda)}}\,.
   \label{E:Deltav}
\end{equation}
Equation (\ref{E:T1}) claims that an applied field produces two distinct
effects on $T_1^{-1}$, one of which originates from the Zeeman energy and
appears in $\bar{n}_k^\sigma$, while the other of which appears via the
nuclear spins giving the characteristic weight
$(v^2k^2+v\hbar\omega_{\rm N})^{-1/2}$ to the electronic transition rate
$\bar{n}_k^\sigma(\bar{n}_k^\sigma+1)$.
The field effect on the nuclear spins escapes observation in critical spin
chains with a linear dispersion at small momenta.
The prefactor $(v^2k^2+v\hbar\omega_{\rm N})^{-1/2}$
is the consequence of the nature of the delta function,
\begin{equation}
   \delta[f(x)]=\sum_i\frac{\delta(x-x_i)}{|f'(x_i)|}\,,
\end{equation}
where $x_i$ is a zero point of an arbitrary regular function $f(x)$, and
therefore generally arises from quadratic dispersion relations of the
relevant electronic excitations, which are the case with ferromagnets
\cite{F433,Y842,Y2324} as well as spin-gapped antiferromagnets.

\begin{figure}
\centerline
{\mbox{\psfig{figure=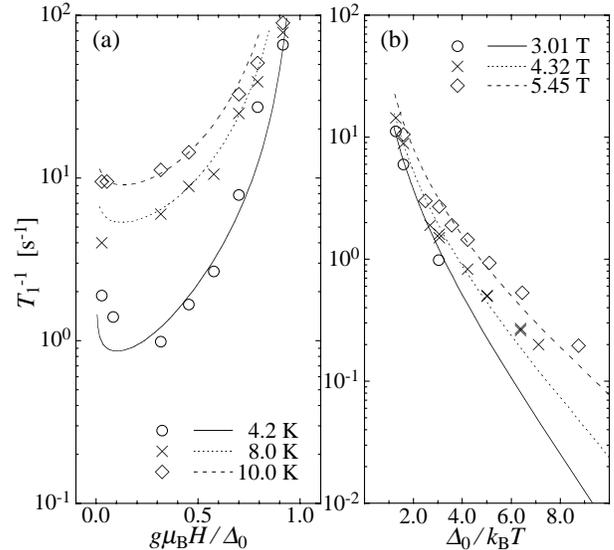,width=80mm,angle=0}}}
\vspace*{1mm}
\caption{Dependences of the nuclear spin-lattice relaxation rate on a
         field parallel to the chain (a) and temperature (b) in NENP
         (symbols) [44] are compared with the modified spin-wave
         calculations (lines), where the field and temperature are scaled
         by the lowest excitation gap ${\mit\Delta}_0$.}
         %[44]=\cite{F11860}
\label{F:T1}
\end{figure}

   Let us fit eq. (\ref{E:T1}) for the proton spin-lattice relaxation-time
measurements on NENP \cite{F11860}.
Although ${\mit\Delta}$ and $v$, given in eq. (\ref{E:Deltav}), depend
on temperature in principle, here we fix
$(v^2k^2+v\hbar\omega_{\rm N})^{-1/2}$ to its zero-temperature value in
the integration (\ref{E:T1}), which is well justified for
$k_{\rm B}T\alt{\mit\Delta}-g\mu_{\rm B}H$ and allows us to inquire
further into eq. (\ref{E:T1}) analytically.
We compare the calculations with the observations in Fig. \ref{F:T1}.
Assuming that $g=2$ and $J/k_{\rm B}=55\,\mbox{K}$ \cite{F11860}, we have
set $A$ equal to $0.024\,\mbox{\AA}^{-3}$, which suggests the distance
between the interacting proton and electron spins being about
$3.5\,\mbox{\AA}$ and is consistent very well with the structural analysis
\cite{M1729}.
Under the present parametrization, the lowest excitation gap is given by
${\mit\Delta}(T=0)/k_{\rm B}\equiv{\mit\Delta}_0/k_{\rm B}
 \simeq 4.0\,\mbox{K}$, which is somewhat smaller than that of NENP,
$12.8\,\mbox{K}$ \cite{F11860}.
However, the scaled function ${\mit\Delta}(T)/{\mit\Delta}_0$ well
reproduces the upward behavior of the Haldane-gap mode as a function of
temperature \cite{Y769}.

\begin{figure}
\centerline
{\mbox{\psfig{figure=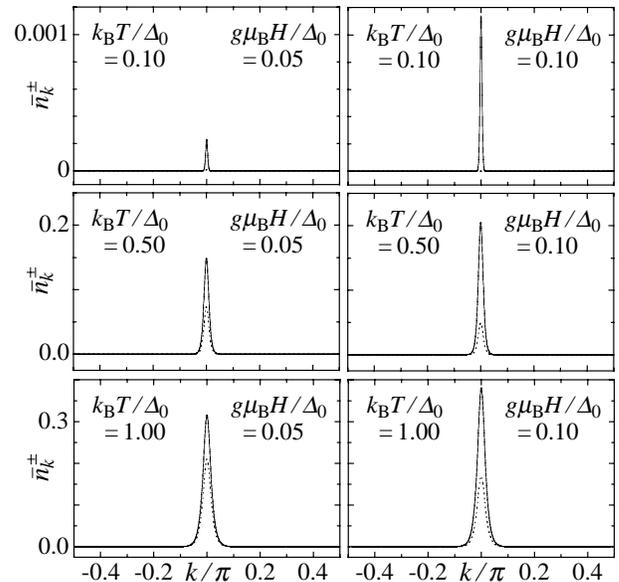,width=80mm,angle=0}}}
\vspace*{1mm}
\caption{The momentum distribution functions $\bar{n}_k^+$ (dotted lines)
         and $\bar{n}_k^-$ (solid lines).}
\label{F:nk}
\end{figure}

   With increasing field, the relaxation rate first decreases moderately
and then increases much more rapidly, at low temperatures in particular.
Although we cannot calculate beyond the critical field
$H_{\rm c}\equiv{\mit\Delta}_0/g\mu_{\rm B}\simeq 9.5\,\mbox{T}$ on the
present formulation, our theory well reproduces the observations
over a wide field range.
For $k_{\rm B}T\alt{\mit\Delta}-g\mu_{\rm B}H$,
the distribution functions may be approximated as
$\bar{n}_k^\pm(\bar{n}_k^\pm+1)\simeq{\rm e}^{-\omega_k^\pm/k_{\rm B}T}$
and therefore eq. (\ref{E:T1}) can further be calculated as
\begin{equation}
   \frac{1}{T_1}\simeq
    \frac{8(g\mu_{\rm B}\hbar\gamma_{\rm N}A)^2}{\pi\hbar v}
    {\rm e}^{-{\mit\Delta}_0/k_{\rm B}T}
    {\rm cosh}\Bigl(\frac{g\mu_{\rm B}H}{k_{\rm B}T}\Bigr)
    K_0\Bigl(\frac{\hbar\omega_{\rm N}}{2k_{\rm B}T}\Bigr)\,,
   \label{E:T1ap}
\end{equation}
where $K_0$ is the modified Bessel function of the second kind.
Provided
$\hbar\omega_{\rm N}\ll k_{\rm B}T\alt{\mit\Delta}-g\mu_{\rm B}H$,
we further reach
\begin{eqnarray}
   &&
   \frac{1}{T_1}\simeq
    \frac{8(g\mu_{\rm B}\hbar\gamma_{\rm N}A)^2}{\pi\hbar v}
    {\rm e}^{-{\mit\Delta}_0/k_{\rm B}T}
    {\rm cosh}\Bigl(\frac{g\mu_{\rm B}H}{k_{\rm B}T}\Bigr)
   \nonumber \\
   &&\qquad\times
   \left[
    0.80908-{\rm ln}\Bigl(\frac{\hbar\omega_{\rm N}}{k_{\rm B}T}\Bigr)
   \right]\,.
\end{eqnarray}
Thus, {\it as $H$ increases, $T_1^{-1}$ should initially decrease
logarithmically and then increase exponentially}.
Figure \ref{F:nk} suggests that the initial decrease of $T_1^{-1}$ turns
from $T_1^{-1}\propto-{\rm ln}H$ to $T_1^{-1}\propto H^{-1/2}$ with
decreasing temperature.
The momentum distribution functions $\bar{n}_k^\pm$ are peaked at $k=0$
and their peaks are sharpened as $T$ decreases.
$\bar{n}_k^\pm$ behaves as $\delta(k)$ in the low-temperature limit.
When we replace $\bar{n}_k^\pm$ by $\delta(k)$, eq. (\ref{E:T1}) gives a
$H^{-1/2}$-linear field dependence of $T_1^{-1}$.

   The temperature dependence is mainly described by the term
${\rm e}^{-({\mit\Delta}_0-g\mu_{\rm B}H)/k_{\rm B}T}$ but further
decorated due to the temperature-dependent energy spectrum.
The inelastic-neutron-scattering peak position of the lowest-energy mode
exhibits an upward behavior with increasing temperature, for
$k_{\rm B}T\agt{\mit\Delta}_0/2$ in particular \cite{R3538,T4677,S3025},
where the slope of ${\rm ln}T_1^{-1}$ to $T^{-1}$ correspondingly
increases with increasing temperature.

   The nuclear magnetic relaxation in the Haldane-gap antiferromagnet NENP
has been interpreted in terms of a modified spin-wave theory.
The field dependence of $T_1^{-1}$ was analyzed in detail and {\it the
minimum of $T_1^{-1}$ as a function of $H$, pending for the past decade,
was solved}.
We consider that such a field dependence of $T_1^{-1}$ is qualitatively
common to spin-gapped antiferromagnets.
We encourage low-temperature $T_1$ measurements on related materials such
as the ferromagnetic-antiferromagnetic bond-alternating compound
(CH$_3$)$_2$CHNH$_3$CuCl$_3$ \cite{M564} and the two-leg ladder
antiferromagnet SrCu$_2$O$_3$ \cite{A3463}.

   The authors are grateful to Dr. N. Fujiwara and Prof. T. Goto for
fruitful discussion.
This work was supported by the Ministry of Education, Culture, Sports,
Science and Technology of Japan, and the Nissan Science Foundation.

\widetext
\end{document}